\documentclass[12pt]{article}
\catcode`\@=11

\@addtoreset{equation}{section}
\def\theequation{\arabic{section}.\arabic{equation}}
     
\catcode`\@=11
\def\thesection{\arabic{section}.}

\def\appendix{\setcounter{section}{0}
        \def\thesection{Appendix \Alph{section}.}
        \def\theequation{\Alph{section}.\arabic{equation}}}
\def\section{\@startsection{section}{1}{\z@}{3.5ex plus 1ex minus
   .2ex}{2.3ex plus .2ex}{\large\bf}}
     
\def\eqnarray{\let\@currentlabel=\theequation\refstepcounter{equation}
    \global\@eqnswtrue
    \global\@eqcnt\z@\tabskip\@centering\let\\=\@eqncr
    $$\halign to \displaywidth\bgroup\@eqnsel\hskip\@centering
      $\displaystyle\tabskip\z@{##}$&\global\@eqcnt\@ne 
       \hfil${{}##{}}$\hfil
      &\global\@eqcnt\tw@ $\displaystyle\tabskip\z@{##}$\hfil 
       \tabskip\@centering&\llap{##}\tabskip\z@\cr}
\def\lefteqn#1{\hbox to 4\arraycolsep{$\displaystyle #1$\hss}}
\long\def\@makefntext#1{\parindent 0cm\noindent
\hbox to 1em{\hss$^{\@thefnmark}$}#1}

\def\IR{{\hbox{{\rm I}\kern-.2em\hbox{\bf R}}}}
\def\IH{{\hbox{{\rm I}\kern-.2em\hbox{\bf H}}}}
\def\IC{{\ \hbox{{\rm I}\kern-.6em\hbox{\bf C}}}}
\def\IZ{{\hbox{{\rm Z}\kern-.4em\hbox{\bf Z}}}}

\def\Tr{{\mathit{Tr}}}
\newcommand{\beq}{\begin{equation}}
\newcommand{\eeq}{\end{equation}}

\begin{document}

%
%
%
%
\def\citen#1{%
\edef\@tempa{\@ignspaftercomma,#1, \@end, }
\edef\@tempa{\expandafter\@ignendcommas\@tempa\@end}%
\if@filesw \immediate \write \@auxout {\string \citation {\@tempa}}\fi
\@tempcntb\m@ne \let\@h@ld\relax \let\@citea\@empty
\@for \@citeb:=\@tempa\do {\@cmpresscites}%
\@h@ld}
%
\def\@ignspaftercomma#1, {\ifx\@end#1\@empty\else
   #1,\expandafter\@ignspaftercomma\fi}
\def\@ignendcommas,#1,\@end{#1}
%
%
\def\@cmpresscites{%
 \expandafter\let \expandafter\@B@citeB \csname b@\@citeb \endcsname
 \ifx\@B@citeB\relax 
    \@h@ld\@citea\@tempcntb\m@ne{\bf ?}%
    \@warning {Citation `\@citeb ' on page \thepage \space undefined}%
 \else
    \@tempcnta\@tempcntb \advance\@tempcnta\@ne
    \setbox\z@\hbox\bgroup 
    \ifnum\z@<0\@B@citeB \relax
       \egroup \@tempcntb\@B@citeB \relax
       \else \egroup \@tempcntb\m@ne \fi
    \ifnum\@tempcnta=\@tempcntb 
       \ifx\@h@ld\relax 
          \edef \@h@ld{\@citea\@B@citeB}%
       \else 
          \edef\@h@ld{\hbox{--}\penalty\@highpenalty \@B@citeB}%
       \fi
    \else   
       \@h@ld \@citea \@B@citeB \let\@h@ld\relax
 \fi\fi%
 \let\@citea\@citepunct
}
%
\def\@citepunct{,\penalty\@highpenalty\hskip.13em plus.1em minus.1em}%
%
%
\def\@citex[#1]#2{\@cite{\citen{#2}}{#1}}%
%
%
\def\@cite#1#2{\leavevmode\unskip
  \ifnum\lastpenalty=\z@ \penalty\@highpenalty \fi 
  \ [{\multiply\@highpenalty 3 #1
      \if@tempswa,\penalty\@highpenalty\ #2\fi 
    }]\spacefactor\@m}
\let\nocitecount\relax  
%
\begin{titlepage}
\vspace{.5in}
\begin{flushright}
UCD-00-11\\
gr-qc/0005017\\
May 2000\\
\end{flushright}
\vspace{.5in}
\begin{center}
{\Large\bf
  Logarithmic Corrections to Black Hole Entropy\\[.6ex]
  from the Cardy Formula}\\
\vspace{.4in}
{S.~C{\sc arlip}\footnote{\it email: carlip@dirac.ucdavis.edu}\\
       {\small\it Department of Physics}\\
       {\small\it University of California}\\
       {\small\it Davis, CA 95616}\\{\small\it USA}}
\end{center}

\vspace{.5in}
\begin{center}
{\large\bf Abstract}
\end{center}
\begin{center}
\begin{minipage}{5in}
{\small
Many recent attempts to calculate black hole entropy from first 
principles rely on conformal field theory techniques.  By examining the 
logarithmic corrections to the Cardy formula, I compute the first-order 
quantum correction to the Bekenstein-Hawking entropy in several 
models, including those based on asymptotic symmetries, horizon 
symmetries, and certain string theories.  Despite very different physical 
assumptions, these models all give a correction proportional to 
the logarithm of the horizon size, and agree qualitatively with recent 
results from ``quantum geometry'' in 3+1 dimensions.   There are some
indications that even the coefficient of the correction may be universal,
up to differences that depend on the treatment of angular momentum
and conserved charges.
}
\end{minipage}
\end{center}
\end{titlepage}
\addtocounter{footnote}{-1}

\section{Introduction}

Recent progress in quantum gravity has made it possible for the first time
to directly compute the Bekenstein-Hawking entropy of a black hole by
counting microscopic states.  Such calculations have been performed both
in string theory \cite{string} and in ``quantum geometry'' \cite{Ashtekar},
and while significant questions remain in each approach, the outlook seems
promising.  The agreement between these very different approaches, and 
the agreement of both with semiclassical calculations that know nothing 
of the details of quantum gravity, has suggested that the asymptotic 
behavior of the density of states may be determined by some simple 
universal feature, such as the algebra of diffeomorphisms at the horizon
\cite{Carlip,Carlip1,Carlip2,Solo}.

As Kaul and Majumdar have recently stressed \cite{Kaul}, however,
different quantum theories of gravity may lead to different higher order
corrections to the Bekenstein-Hawking entropy.  These corrections may
display differences, or---less probably but more interestingly---relations
among quantizations.  In Ref.\ \cite{Kaul}, Kaul and Majumdar compute
the lowest order corrections to the Bekenstein-Hawking entropy in a
particular formulation \cite{Kaul2} of the ``quantum geometry'' program
of Ashtekar et al.  They find that the leading correction is logarithmic, with
\beq
S \sim {A\over4G} - {3\over2}\ln\left( {A\over4G}\right) + 
  \hbox{\it const.} + \dots
\label{a1}
\eeq

Although the existing computations of black hole entropy have very
different physical starting points, most use techniques from
two-dimensional conformal field theory at an intermediate stage.  
This may be no more than a useful trick---the Cardy formula 
\cite{Cardy,Cardy2} makes it particularly easy to count states in a 
two-dimensional conformal field theory---but there are suggestions 
\cite{Maldacena,Larsen} that such conformal field theories 
really provide a universal description of low-energy black hole 
thermodynamics.  Whatever the origin of the conformal symmetry, 
however, the same trick that allows us to determine the asymptotic 
density of states also permits a direct computation of the leading
quantum corrections to the Bekenstein-Hawking entropy.

In this paper, I compute the logarithmic terms in the Cardy formula and 
use the results to obtain quantum corrections to black hole entropy.  I 
examine a number of approaches, including Strominger's asymptotic 
symmetry analysis \cite{Strominger} of the  (2+1)-dimensional BTZ 
black hole \cite{BTZ}; the  methods of Ref.\ \cite{Carlip,Carlip1,Carlip2}, 
which are based on the behavior of symmetries at the black hole horizon 
in any dimension; and the string theoretical counting of D-brane states 
for BPS black holes \cite{Vafa,Horowitz}.  In all cases, I find qualitative 
agreement with the ``quantum geometry'' result (\ref{a1}), with answers 
that differ by a factor of two in the coefficient of the logarithm and 
(sometimes) an additional term that depends on conserved charges.  
I suggest that these differences may be traced back to an ambiguity in 
the treatment of angular momentum and other conserved quantities, 
which may lead to the counting of different sets of states.  If this is the 
case, these results represent a surprising new universality in the 
logarithmic corrections to the Bekenstein-Hawking entropy.

\section{Logarithmic Corrections to the Cardy Formula \label{secb}}

The recent ``first principles'' computations of black hole entropy, 
whatever their physical starting point, typically rely at some critical
stage on the Cardy formula \cite{Cardy,Cardy2} for the density of 
states in a two-dimensional conformal field theory.  I will start by 
reviewing the derivation of this formula, in order to obtain logarithmic 
corrections to the density of states.

Let us begin with an arbitrary two-dimensional conformal field theory 
with central charge $c$, with the standard Virasoro algebra
\begin{eqnarray}
\left[L_m,L_n\right] &=& 
     (m-n)L_{m+n} + {c\over12}m(m^2-1)\delta_{m+n,0}
\nonumber\\ 
\left[{\bar L}_m,{\bar L}_n\right] &=& (m-n){\bar L}_{m+n} 
    + {c\over12}m(m^2-1)\delta_{m+n,0}\\
\left[L_m,{\bar L}_n\right] &=& 0 \nonumber
\label{b1}
\end{eqnarray}
for the generators $L_n$, ${\bar L}_n$ of holomorphic and
antiholomorphic diffeomorphisms.  The partition function on the 
two-torus of modulus $\tau = \tau_1 + i\tau_2$ is defined to be
\beq
Z(\tau,\bar\tau) =
\Tr\ e^{2\pi i\tau L_0}e^{-2\pi i{\bar\tau}{\bar L}_0} =
\sum\rho(\Delta,{\bar \Delta})
e^{2\pi i\Delta\tau}e^{-2\pi i{\bar \Delta}{\bar\tau}} .
\label{b2}
\eeq
For a unitary theory, $\rho$ is the number of states with eigenvalues
$L_0 = \Delta$, ${\bar L}_0 = {\bar\Delta}$, as can be seen by inserting
a complete set of states into the trace.  For a nonunitary theory, $\rho$
is the difference between the number of positive- and negative-norm
states with appropriate eigenvalues.

If we could somehow determine the partition function, we could extract 
the density of states by contour integration.  Treat $\tau$ and $\bar\tau$ 
as independent complex variables (this is not necessary, but it simplifies 
the computation), and let $q=e^{2\pi i\tau}$ and ${\bar q}=
e^{2\pi i{\bar\tau}}$.  Then
\beq
\rho(\Delta,{\bar \Delta}) =
{1\over(2\pi i)^2} \int{dq\over q^{\Delta+1}}
{d{\bar q}\over {\bar q}^{{\bar \Delta}+1}} Z(q,{\bar q}) ,
\label{b3}
\eeq
where the integrals are along contours that enclose $q=0$ and ${\bar q}=0$.
Of course, it is rare that we actually know $Z(q,{\bar q})$.  But Cardy has 
shown that it is still possible to relate the behavior of the partition function 
at high ``energy'' to its simpler behavior at low ``energy,''  thus giving us 
some control over the integral (\ref{b3}).

Cardy's basic result \cite{Cardy,Cardy2} is that the quantity
\beq
\Tr\ e^{2\pi i(L_0-{c\over24})\tau}
  e^{-2\pi i({\bar L}_0-{c\over24}){\bar\tau}}
= e^{{\pi c\over6}\tau_2}Z(\tau,\bar\tau)
\label{b4}
\eeq
is modular invariant, and in particular invariant under the large 
diffeomorphism $\tau\rightarrow-1/\tau$ that interchanges the
circumferences of the torus.  The argument is universal, involving only 
some general properties of conformal field theory.  We can use this 
result to attempt to evaluate the integral (\ref{b3}) by steepest descent.  
To do so, let $\Delta_0$ be the lowest eigenvalue of $L_0$ (often but not 
always zero), and define
\beq
{\tilde Z}(\tau) = \sum \rho(\Delta)e^{2\pi i(\Delta-\Delta_0)\tau} =
\rho(\Delta_0) + \rho(\Delta_1)e^{2\pi i(\Delta_1-\Delta_0)\tau} + \dots
\label{b5}
\eeq
(For simplicity, I have suppressed the $\bar\tau$ dependence.)  It is then 
straightforward to show\footnote{See \cite{Carlip3} for details.  A closely
related but more general result, obtained from a sophisticated number
theoretical analysis that yields a much more complete description of 
the asymptotics, is discussed in \cite{Moore}.} that
\beq
\rho(\Delta) = \int d\tau\, e^{-2\pi i\Delta\tau}
e^{-2\pi i\Delta_0{1\over\tau}}
e^{{2\pi ic\over24}\tau}e^{{2\pi ic\over24}{1\over\tau}}
{\tilde Z}(-{1/\tau}) .
\label{b6}
\eeq
By construction, ${\tilde Z}(-{1/\tau})$ approaches a constant, 
$\rho(\Delta_0)$, for large $\tau_2$, so the integral (\ref{b6}) can safely 
be evaluated by steepest descents provided that the imaginary part of 
$\tau$ is large at the saddle point.

The integral we need has the form
\beq
I[a,b] = \int d\tau\, e^{2\pi ia\tau + {2\pi i b\over\tau}}f(\tau) .
\label{b7}
\eeq
The argument of the exponent is extremal at $\tau_0 = \sqrt{b/a}$, and
expanding around this extremum, we find
\beq
I[a,b] \approx \int d\tau\, 
  e^{4\pi i\sqrt{ab} + {2\pi ib\over\tau_0^3}(\tau-\tau_0)^2}f(\tau_0)
  = \left(-{b\over4a^3}\right)^{1/4}e^{4\pi i\sqrt{ab}} f(\tau_0) .
\label{b8}
\eeq
In particular, if $\Delta_0$ is small ($\Delta_0\ll c$) and $\Delta$ is large, 
the integral (\ref{b6}) yields
\beq
\rho(\Delta) \approx \left({c\over96\Delta^3}\right)^{1/4}
   \exp\left\{ 2\pi\sqrt{c\Delta\over6}\right\} .
\label{b9}
\eeq
The exponential term in (\ref{b9}) gives the standard Cardy formula, 
but we have now found the leading correction as well.

We must next ask how reliable this approximation is.  For $f(\tau)$ 
constant, the integral (\ref{b7}) can be performed explicitly, yielding a 
Bessel function, whose asymptotic behavior agrees with (\ref{b8}) with 
additional terms that are exponentially suppressed.  Corrections from 
the nonconstancy of $f(\tau)$ may also be computed, and for $f(\tau) =
{\tilde Z}(-1/\tau)$, it is easy to check that these are again exponentially 
suppressed.  For large $\Delta$, the expression (\ref{b9}) thus gives a 
reliable first-order correction to the standard Cardy formula.

It should be stressed that the central charge $c$ appearing in (\ref{b9}) 
is the full central charge of the conformal field theory.  In general, 
$c$ will consist of a ``classical'' term $c_{\hbox{\scriptsize\it class}}$, 
which already appears in the Poisson brackets of the $L_n$, plus a 
quantum correction that can change the exponent in (\ref{b9}) from 
its ``classical'' value.  In contrast to the prefactor in (\ref{b9}), this 
correction is likely to be highly model-dependent.  Nevertheless, a bit 
can be said about its general features; see Appendix B.

\section{The BTZ Black Hole}

As our first application of Eqn.\ (\ref{b9}), let us evaluate the logarithmic
corrections to Strominger's derivation of the entropy of the BTZ black
hole.  This (2+1)-dimensional black hole has a metric
\beq
ds^2 = -N^2dt^2 + N^{-2}dr^2 + r^2\left( d\phi + N^\phi dt\right)^2
\label{c1}
\eeq
with
\beq
N = \left(-8GM + {r^2\over\ell^2} + {16G^2J^2\over r^2}\right)^{1/2} , 
\qquad N^\phi = - {4GJ\over r^2} \qquad  (|J|\le M\ell) ,
\label{c2}
\eeq
and solves the vacuum Einstein field equations with a cosmological constant 
$\Lambda = -1/\ell^2$.  The spacetime is thus asymptotically anti-de Sitter,
and has an outer (event) and an inner horizon at
\beq
r_\pm{}^2={4GM\ell^2}\left \{ 1 \pm
\left [ 1 - \left({J\over M\ell}\right )^2\right ]^{1/2}\right \} ,
\label{c3}
\eeq
i.e.,
\beq
M={r_+{}^2+r_-{}^2\over8G\ell^2}, \quad J={r_+ r_-\over4G\ell} \ .
\label{c4}
\eeq

As Brown and Henneaux first noted \cite{BrownH}, the asymptotic 
symmetries of Einstein gravity in 2+1 dimensions with negative 
$\Lambda$ are described by a pair of Virasoro algebras, with central 
charges
\beq
c = {\bar c} =  {3\ell\over2G} .
\label{c5}
\eeq
Intuitively, these are the symmetries of the adS ``cylinder at infinity,'' 
obtained by restricting diffeomorphisms in the bulk to those that preserve 
adS boundary conditions.  The central charges (\ref{c5}) are classical, but 
they will presumably be reflected in any quantum theory of gravity.  Thus
in any quantum theory of gravity that has the correct classical limit, the 
fields should transform under representations of these Virasoro algebras.
Given some plausible assumptions \cite{Carlip3}---for example, that 
$\Delta_0$ is small---one should therefore be able to use the Cardy 
formula to compute the asymptotic density of states.

Now, the generators of the Brown-Henneaux Virasoro algebras can be
computed explicitly: they are simply the Hamiltonian and momentum
constraints of general relativity smeared against appropriate vector fields.
For the BTZ black hole, one finds that up to an ambiguous additive 
constant \cite{Banados},
\beq
\Delta = {(r_+ + r_-)^2\over16G\ell} , \qquad 
{\bar\Delta} = {(r_+ - r_-)^2\over16G\ell} .
\label{c6a}
\eeq
As Strominger observed, Eqns.\  (\ref{c5}) and (\ref{c6a}) can be used to 
evaluate the exponent in (\ref{b9}), yielding
\beq
2\pi\sqrt{c\Delta\over6} + 2\pi\sqrt{{\bar c}{\bar\Delta}\over6} = 
{2\pi r_+\over4G} ,
\label{c7}
\eeq
giving the standard Bekenstein-Hawking entropy for the 
(2+1)-dimensional black hole.

It is now easy to read off the logarithmic corrections to the entropy.  From
(\ref{b9}),
\beq
\rho(\Delta,{\bar\Delta}) \approx  {8G\ell^2\over(r_+^2 - r_-^2)^{3/2}}
     \exp\left\{{2\pi r_+\over4G}\right\} .
\label{c8}
\eeq
Thus
\beq
S \sim {2\pi r_+\over4G} -  {3\over2}\ln\left({r_+^2-r_-^2\over G^2}\right) 
   + \hbox{\it const.} = {2\pi r_+\over4G}  
   - {3\over2}\ln{2\pi r_+\over G} - {3\over2}\ln{\kappa\ell} + \hbox{\it const.}
\label{c9}
\eeq
where
\beq
\kappa = {r_+^2-r_-^2\over\ell^2r_+}
\label{c9a}
\eeq
is the surface gravity.  The logarithmic terms in (\ref{c9}) have the same form
as those found by Kaul and Majumdar for the nonrotating (3+1)-dimensional
black hole.  Because of the term involving $\kappa$, however, the coefficients
are different; in particular, if one restricts to zero angular momentum ($r_-=0$), 
one finds logarithmic term that differs from (\ref{a1}) by a factor 
of two.

There is an alternative derivation of the BTZ black hole entropy, first
proposed  in Ref.\ \cite{Carlip4}, that directly counts states of an induced
$\hbox{SL}(2,{\IR})\times\hbox{SL}(2,{\IR})$ Wess-Zumino-Witten
model at the black hole horizon.  While the horizon radius $r_+$ has a
natural expression in such a WZW model, it is difficult to fix $r_-$, so 
one instead fixes a conjugate variable (essentially a component of the
triad at the horizon).  As explained in Appendix B of Ref.\ \cite{Carlip3},
the resulting partition function can be viewed as a functional Fourier
transform of the partition function (\ref{b2}), where the central charge 
in (\ref{b2}) for an $\hbox{SL}(2,{\IR})$ WZW model is $c\approx3$.  
One finds
\beq
Z = \sum_N 
\rho(N)\exp\left\{2\pi i\tau \left( N - {k^2r_+^2\over\ell^2}\right)\right\}
\label{c10}
\eeq
with
\beq
\rho(N) \sim \sum_{n=0}^N \rho_0(N)\rho_0(N-n) ,
\label{c11}
\eeq
where $\rho_0$ is the partition function for an $\hbox{SL}(2,{\IR})$ 
WZW model and $k=\ell/4G$.  

In the large $k$  limit, the three oscillators of $\hbox{SL}(2,{\IR})$ can 
be treated independently, and (\ref{b9}) gives
\beq
\rho(N) \sim \sum_{n=0}^N N^{-3/4}(N-n)^{-3/4}
   \exp\left\{ \sqrt{2}\pi \left( \sqrt{n} + \sqrt{N-n} \right)\right\} .
\label{c12}
\eeq
We can evaluate this expression by approximating the sum as an integral
and using the method of steepest descents, obtaining
\beq
\rho(N) \sim N^{-3/4}e^{2\pi\sqrt{N}} .
\label{c13}
\eeq
In the formalism of Ref.\ \cite{Carlip4}, the partition function 
(\ref{c10}) is subject to a physical state condition $\Delta=0$, that is, 
$N = k^2r_+^2/\ell^2$.  We thus obtain a density of states
\beq
\rho(N) \sim (r_+/G)^{-3/2} \exp\left\{2\pi r_+\over4G\right\} .
\label{c14}
\eeq
The resulting logarithmic correction to the entropy agrees exactly
with that of Kaul and Majumdar.

It is interesting to note that logarithmic corrections of this sort are 
absent in the Euclidean path integral approach to BTZ black hole entropy.  
The first-order corrections were calculated in that formalism in Ref.\ 
\cite{CarTeit}; they give an exponentially suppressed contribution to
the density of states, with no power law prefactor that would translate
into a logarithmic correction to the entropy.

\section{Horizon Conformal Field Theory}

The conformal field theory derivations of the preceding section rely 
on special features of the (2+1)-dimensional black hole.  The results
are more general than they might appear at first sight, since many 
black holes in string theory have a near-horizon structure that looks 
like that of a BTZ black hole \cite{string,Carlip3}.  Others do not, 
however, and the methods do not directly generalize directly to higher 
dimensions.

A different conformal field theory approach to black hole entropy 
has recently been proposed, based on a possible universal Virasoro
algebra at the horizon \cite{Carlip,Carlip1,Carlip2}.  This algebra
is obtained by treating the horizon as a boundary and considering the
behavior of the algebra of diffeomorphisms of the ``$r$--$t$ plane''
near the horizon.  The proper choice of boundary conditions is not
entirely clear, but several different choices give rise to a Virasoro
algebra with central charge
\beq
c = {3A\over2\pi G}{\beta\over\kappa}
\label{d1}
\eeq
and an $L_0$ eigenvalue
\beq
\Delta = {A\over16\pi G}{\kappa\over\beta} ,
\label{d2}
\eeq
where $A$ is the horizon area (in any dimension), $\kappa$ is the
surface gravity, and $\beta$ is an undetermined periodicity.  An
analysis of the Liouville theory near the horizon obtained from
dimensional reduction of Einstein gravity gives a similar result
\cite{Solo}.

It is easy to see that these values of $c$ and $\Delta$, inserted into
the Cardy formula, give the standard Bekenstein-Hawking entropy.
But we can now go further, and compute the logarithmic corrections:
Eqn.\ (\ref{b9}) yields
\beq
\rho(\Delta) \sim {c\over12}\left({A\over8\pi G}\right)^{-3/2}
   \exp\left\{ {A\over4G} \right\} .
\label{d3}
\eeq
If we can now choose $\beta$ to be such that $c$ is a universal constant, 
independent of $A$, we find an entropy
\beq
S \sim {A\over4G} - {3\over2}\ln\left( {A\over4G}\right) + 
\hbox{\it const.} + \dots ,
\label{d4}
\eeq
in agreement with the result (\ref{a1}) of Kaul and Majumdar.

\section{String Theory}

Much of the current interest in black hole entropy was sparked by the
discovery by Strominger and Vafa \cite{Vafa} that for extremal (BPS)
black holes in string theory, one could compute the Bekenstein-Hawking
entropy by counting D-brane states.  The relevant configurations are
obtained by compactifying a suitable string theory on a manifold with 
the topology $M\times S^1$---$M$ is $K3$ in the case considered in 
Ref.\ \cite{Vafa}, but may be different for other black holes---and 
considering a collection of D-branes wrapped around cycles of 
$M\times S^1$.  To count states, one takes the radius of the $S^1$ 
factor to be large compared to $M$, and describes the low-energy 
excitations of the D-branes in terms of fields moving on $S^1$.  This 
description involves a weak-coupling approximation, and is unrealistic 
for a true black hole.  For BPS configurations, however, one can argue 
that the density of states is protected by nonrenormalization theorems 
when the coupling is increased.

The sigma model describing the excitations on $S^1$ is a two-dimensional 
conformal field theory, albeit a conformal field theory very different from 
those considered in the preceding sections of this paper.  Thus Cardy's 
formula may be used to count states, and we may again appeal to Eqn.\ 
(\ref{b9}) to find the logarithmic corrections to the entropy.

For the five-dimensional black holes investigated by Strominger and Vafa,
the relevant conformal field theory has central charge
\beq
c \approx 3Q_F^2
\label{e1}
\eeq
and $L_0$ eigenvalue
\beq
\Delta = Q_H ,
\label{e2}
\eeq
where $Q_F$ is the Ramond-Ramond charge and $Q_H$ is the momentum 
around the $S^1$.  These quantities translate into charges of the associated
black holes, and the entropy obtained from the exponential term in the Cardy
formula turns out to be $A/4G$, where the horizon area is
\beq
A = 8\pi\sqrt{{Q_HQ_F^2\over2}} .
\label{e3}
\eeq
By (\ref{b9}), the leading correction to the entropy is thus
\beq
S \sim {A\over4G} - {3\over2}\ln\left( {A\over4G}\right) +  2\ln Q_F +
  \hbox{\it const.} + \dots
\label{e4}
\eeq
We again obtain a logarithmic correction identical to that of Kaul and 
Majumdar, along with an extra term depending on the Ramond-Ramond 
charge.

A similar analysis can be performed for the four-dimensional black holes 
of Horowitz et al.\ \cite{Horowitz,Horowitz2}, which are obtained from a 
string theory compactified on a six-torus, with charges $Q_2$, $Q_5$, 
$Q_6$, and $n$ carried by two-branes, five-branes, six-branes, and strings 
wrapped around cycles of the torus.  In the extremal limit, one obtains a 
conformal field theory with
\beq
c = 6Q_2Q_5Q_6, \qquad \Delta = n ,
\label{e5}
\eeq
corresponding to a black hole with horizon area
\beq
A = 2\pi\sqrt{Q_2Q_5Q_6n} .
\label{e6}
\eeq
The leading correction to the entropy is thus
\beq
S \sim {A\over4G} - {3\over2}\ln\left( {A\over4G}\right) 
   + \ln(Q_2Q_5Q_6) +  \hbox{\it const.} + \dots
\label{e7}
\eeq
Again, we obtain an area term of the form (\ref{a1}), plus corrections that 
depend on conserved charges.

Larsen has proposed a related conformal field theoretical picture for a large
class of five-dimensional charged, rotating black holes \cite{Larsen,Kastor}.  
Such black holes are characterized by a mass $M$, three conserved charges
$Q_i$, and two angular momenta $J_{R,L}$.  The mass and charges can be
parametrized as
\beq
M = {1\over2}\mu\sum_{i=1}^3\cosh2\delta_i , \qquad
Q_i = {1\over2}\sinh2\delta_i .
\label{e8}
\eeq
Larsen shows that many of the properties of such black holes can be understood
in terms of a left- and right-moving conformal field theory, each with central 
charge $c=6$, with $L_0$ eigenvalues
\beq
\Delta_{R,L} = {1\over4}\mu^3
    \left(\prod_i\cosh\delta_i \mp \prod_i\sinh\delta_i\right)^2 - J_{R,L}^2 .
\label{e9}
\eeq
The Bekenstein-Hawking entropy follows from Cardy's formula, and can be 
written in the form
\beq
S_{B-H} = S_R + S_L ,\qquad S_{R,L} = 2\pi\sqrt{{c\Delta_{R,L}\over6}}  .
\label{e10}
\eeq
Since the central charge $c$ is just a number, it is easy to use Eqn.\ 
(\ref{b9}) to obtain the logarithmic corrections to (\ref{e10}).  Using the
results of \cite{Larsen}, we obtain
\beq
S \sim {A_+\over4G} - {3\over2}\ln(S_LS_R) + \hbox{\it const.} + \dots
    = {A_+\over4G} - {3\over2}\ln\left[ \left(A_+\over4G\right)^2 -
    \left(A_-\over4G\right)^2\right] + \hbox{\it const.} + \dots ,
\label{e11}
\eeq
where $A_\pm$ are the areas of the outer and inner horizons.  This result
may be compared to the nearly identical expression (\ref{c9}) for the BTZ 
black hole.

\section{Speculations}

Although the models of black hole entropy considered here involve very
different physical pictures of microscopic states, all use two-dimensional
conformal field theory as a crucial tool.  This makes it possible to compute  
the leading logarithmic corrections to the Bekenstein-Hawking entropy  
in a simple and systematic manner.  The resulting entropy takes the general 
form
\beq
S \sim {A\over4G} - {3\over2}\ln\left( {A\over4G}\right) + \ln F(Q) +
    \hbox{\it const.} + \dots ,
\label{f1}
\eeq
where $F(Q)$ is some function of angular momentum and other conserved 
charges.

The existence of logarithmic corrections of the form $\ln(A/4G)$ is thus 
a general feature of black hole entropies obtained in this manner.  The
interesting question is whether the factor of $-3/2$ in (\ref{f1}), which
also appears in the results of Kaul and Majumdar, is also universal.  

The problem, of course, is that the charges in $F(Q)$ and the horizon area 
are not, in general, independent, so there is some ambiguity in the division 
of the right-hand side of (\ref{f1}) into separate terms.  To explore this 
issue further, let us return to the Cardy formula (\ref{b9}), and note that 
it can be rewritten as
\beq
\ln\rho(\Delta) = S_0 - {3\over2}\ln S_0 + \ln c + \hbox{\it const.} ,
\label{f2}
\eeq
where $S_0 = 2\pi\sqrt{c\Delta/6}$.  Hence if the entropy is obtained from a 
single conformal field theory, and if the central charge $c$ is ``universal'' 
in the sense of being independent of the horizon area, the factor of $-3/2$ 
will always appear.

This sort of ``universality'' of the central charge seems to be a sensible
requirement for any fundamental conformal field theoretical explanation
of black hole entropy: one should surely be looking for a single conformal 
field theory to describe black holes with arbitrary masses and charges.  The 
conformal theories of Refs.\ \cite{Carlip,Carlip1,Carlip2} do not satisfy this 
demand, but they are presumably not yet the final word on horizon boundary
conditions.

The requirement of a single conformal field theory is less clear.  One might
reasonably argue that there should be left-moving and a right-moving
sectors, as there are for the asymptotic symmetries of the BTZ black hole.  
Larsen, for instance, has suggested that these sectors could be associated 
with the inner and outer horizon \cite{Larsen}.  The presence of two
conformal field theories changes the form of the leading correction in 
(\ref{f2}), which becomes
\beq
\ln\rho(\Delta_L) + \ln\rho(\Delta_R) = S_L + S_R - {3\over2}\ln(S_LS_R) 
   + \ln(c_Lc_R) + \hbox{\it const.} 
\label{f3}
\eeq
It is evident that if $S_L + S_R $ gives the standard Bekenstein-Hawking 
entropy, the logarithmic correction is no longer $-{3\over2}\ln S_{B-H}$,
but rather takes a form more like that of Eqn.\ (\ref{e11}).

This difference appears to account for much of the variation among the
logarithmic terms found in this paper.  The difference may be related to
the choice of how to treat angular momentum and other conserved charges
when counting states, and thus implicitly to the choice of which states to 
count.  For the (2+1)-dimensional black hole, for example, Strominger's 
approach starts with a black hole with a fixed angular momentum, and gives 
an expression of the general form (\ref{f3}).  The earlier approach of Ref.\ 
\cite{Carlip4}, on the other hand, does not require a specified angular 
momentum, and yields an expression of the form (\ref{f2}).  It will be 
interesting to see how the results of Kaul and Majumdar \cite{Kaul}, 
which are based on boundary conditions specific to a nonrotating black 
hole, change when more general rotating boundary conditions are 
incorporated.

\vspace{1.5ex}
\begin{flushleft}
\large\bf Acknowledgements
\end{flushleft}
This work was supported in part by Department of Energy grant
DE-FG03-91ER40674.
 
\appendix
\section{The String Level Density and the Cardy Formula}

The expression (\ref{b9}) for the density of states differs from the 
``level density'' of string theory \cite{GSW}, which counts the oscillator 
states of a string theory.  In this appendix I investigate the differences, 
while simultaneously obtaining a useful check of the methods of section 
\ref{secb}

Let us begin by using the procedure that led to Eqn.\ (\ref{b9}) to 
evaluate the ``partition function'' $p(n)$ of Ramanujan and Hardy 
\cite{Ramanujan}, the number of partitions of an integer $n$ into a 
sum of smaller integers.  It is easy to see that $p(n)$ is also the number 
of oscillator states of a bosonic string in one transverse dimension.  The 
generating function for $p(n)$ is
\beq
G(\tau) = \sum p(n)e^{2\pi i n\tau}  
   = \prod_{n=1}^\infty \left( 1 - e^{2\pi in\tau} \right)^{-1}
   = e^{{2\pi i\tau\over24}} \eta^{-1}(\tau) ,
\label{A1}
\eeq
where $\eta(\tau)$ is the Dedekind eta function.  This is almost a chiral
partition function $Z(\tau)$ for a $c=1$ conformal field theory, but not
quite: the modular transformation properties of $\eta(\tau)$ differ slightly 
from the standard conformal field theory form.  Indeed,
\beq
\eta^{-1}(-1/\tau) = (-i\tau)^{-1/2}\eta^{-1}(\tau) ,
\label{A2}
\eeq
so unlike Cardy's partition function (\ref{b4}), the quantity
$e^{-{2\pi i\tau/24}}G(\tau)$ is not exactly modular invariant.  It is 
straightforward to incorporate the extra factor in the transformation
(\ref{A2}) into the integral (\ref{b6}), however: one must simply 
replace ${\tilde Z}(-{1/\tau})$ by $(-i\tau)^{1/2}{\tilde Z}(-{1/\tau})$.  
This translates into a term $f(\tau_0) = (-i\tau_0)^{1/2} = (-b/a)^{1/4}$ 
in Eqn.\ (\ref{b8}), giving
\beq
p(n) \sim {1\over\sqrt{48}n}e^{\pi\sqrt{2n\over3}} ,
\label{A3}
\eeq
in exact agreement with the asymptotic results of Ramanujan and Hardy.

The generalization to a partition with $N$ ``colors'' is immediate.  The
relevant generating function is now
\beq
G_N(\tau) = \sum p_N(n)e^{2\pi i n\tau} 
   = \prod_{n=1}^\infty \left( 1 - e^{2\pi in\tau} \right)^{-N}
   = \left(e^{{2\pi i\tau\over24}} \eta^{-1}(\tau)\right)^N ,
\label{A4}
\eeq
giving a factor of $(-i\tau_0)^{N/2}$ in Eqn.\ (\ref{b8}) and an asymptotic
behavior
\beq
p_N(n) \sim {1\over\sqrt2} \left({N\over24}\right)^{{N+1\over4}}
   n^{-{N+3\over4}}\exp\left\{2\pi\sqrt{{Nn\over6}}\right\} .
\label{A5}
\eeq
This can be recognized as the ``level density'' for a bosonic string in $N$
transverse dimensions, as first computed by Huang and Weinberg in the 
context of the Veneziano model \cite{Weinberg}.  This expression and 
its superstring generalization have been used by Solodukhin to examine
logarithmic corrections in the string-black hole correspondence 
\cite{Solo2}.

Let us now try to understand the reason for the differences between the 
densities of states (\ref{b9}) in conformal field theory and the level
density (\ref{A5}).  The key observation is that the partition function 
for a scalar field in conformal field theory is not actually given by Eqn.\
(\ref{A1}), but is rather \cite{DMS}
\beq
Z(\tau) = 
     \tau_2^{-1/4}\,e^{{2\pi i\tau\over24}} \eta^{-1}(\tau)  ,
\label{A6}
\eeq
where the extra factor of $\tau_2^{-1/4}$ comes from zeta function
regularization of a determinant, that is, from zero-modes of the boson.
Under the transformation $\tau\rightarrow-1/\tau$, we have
$\tau_2\rightarrow\tau_2/|\tau|^2$, and it is easily checked that the 
resulting factor of $|\tau|^2$ is just what is needed to compensate for
the transformation (\ref{A2}) of $\eta(\tau)$, restoring the Cardy formula 
(\ref{b9}).  

The role of the zero-modes can be explored further by considering a
string compactified on a circle of radius $R$.  The Virasoro generators
$L_0$ and ${\bar L}_0$ are then \cite{DMS}
\begin{eqnarray}
L_0 &=& {1\over2}\left( {r\over2R} +  sR\right)^2 + N \nonumber\\
{\bar L}_0 &=& {1\over2}\left( {r\over2R} -  sR\right)^2 +  {\bar N} ,
\label{A7}
\end{eqnarray}
where $r$ and $s$ are integer-valued momentum and winding numbers 
and $N$ and $\bar N$ are the usual oscillator number operators.  The 
level density (\ref{A3}) counts oscillator states alone, implicitly taking
$r=s=0$.  To incorporate the winding states, we should instead consider 
the sum
\beq
\rho(\Delta,{\bar\Delta}) \sim \sum_{r,s} 
   p\left(\Delta - {1\over2}\left( {r\over2R} +  sR\right)^2\right)\cdot
   p\left({\bar\Delta} - {1\over2}\left( {r\over2R} -  sR\right)^2\right) .
\label{A8}
\eeq
Approximating the sums by integrals, we can write (\ref{A8}) in 
terms of modified Bessel functions $I_1(2\pi\sqrt{\Delta/6})$ and
$I_1(2\pi\sqrt{{\bar\Delta}/6})$, and it is straightforward to
check that the asymptotic behavior of (\ref{A8}) is precisely that of
a $c=1$ conformal field theory, as given by Eqn.\ (\ref{b9}).  The Cardy
formula (\ref{b9}) can thus be understood as a total density of states,
including both the oscillator states counted by (\ref{A5}) and the
winding states or zero-modes that can contribute to $\Delta$ and
$\bar\Delta$.

\section{Corrections to the Central Charge}

As noted at the end of section 2, the central charge in Eqn.\ (\ref{b9}) 
generally takes the form
\beq
c = c_{\hbox{\scriptsize\it class}} + \delta c ,
\label{B1}
\eeq
where the ``classical'' central charge $c_{\hbox{\scriptsize\it class}}$ is
already present in the Poisson brackets of the Virasoro generators.  The
quantum correction $\delta c$ can be evaluated in some models, and can
offer an interesting new test of the statistical mechanical picture.  
For certain black holes in string theories compactified on Calabi-Yau 
manifolds, for example, the one-loop contribution to $\delta c$ gives 
rise to an area-independent term in the Cardy formula \cite{MSW,Cardosa}, 
which can be interpreted in Wald's Noether charge formalism \cite{Wald} 
as the entropy due to a curvature-squared term in the effective action.  
Since the coefficient of this term in the action is independently calculable 
in string theory, this agreement provides a delicate test of the statistical 
mechanical formalism.  Similarly, it has been argued that the central 
charge (\ref{c5}) for the BTZ black hole should be shifted to 
\beq
c = {\bar c} =  {3\ell\over2G} + \beta + {4G\over\ell}\gamma 
\label{B2}
\eeq
for constants $\beta$ and $\gamma$ that depend on details of the relevant 
conformal field theory \cite{Behrndt}.  For an appropriate choice of
$\beta$, the resulting correction to the Cardy formula reproduces the 
shift in entropy found in Ref.\ \cite{CarTeit} from a one-loop path 
integral computation.   

There seems to be no reason to expect the quantum corrections 
$\delta c$ to have any special universal properties.  We should 
therefore ask whether the logarithmic corrections to the entropy 
described in this paper are really the leading corrections.  Although
I know of no completely rigorous argument, for large black holes it 
seems quite likely that they are:

Recall first that the Bekenstein-Hawking formula for entropy can
be obtained from quantum field theory in a {\em classical\/} black
hole background \cite{Hawking}.  Such a result will receive quantum 
gravitational corrections, of course, but since the curvature near the
horizon decreases as the horizon area $A$ increases, one 
expects these corrections to be small for large black holes.  Equivalently, 
a term in the four-dimensional effective action with $n$ powers of the 
curvature will have a coupling constant of dimension $[L]^{2n-4}$,
which can multiply at most a factor of $A^{2-n}$.  Quantum 
corrections to the entropy involve $n>1$ powers of the curvature,
and  should thus be suppressed by powers of the horizon area.  Such
an effective action picture may fail for small black holes, but it should 
be approximately valid in the semiclassical regime, that is, for black 
holes whose horizons are large compared to the Planck scale.

Now consider an expansion of $\delta c/c_{\hbox{\scriptsize\it class}}$ 
in powers of $A$.  The leading correction to the entropy will come from
the exponent of Eqn.\ (\ref{b9}), yielding
\beq
S \sim {A\over4\hbar G}\left( 
    1 + {\delta c\over c_{\hbox{\scriptsize\it class}}}\right)^{1/2} .
\label{B3}
\eeq
Positive powers of $A$ in $\delta c/c_{\hbox{\scriptsize\it class}}$
would dominate this expression for large $A$, leading to a breakdown of 
the Bekenstein-Hawking formula in precisely the regime in which it
should be most reliable.  As argued above, such terms are thus unlikely.

The next terms in the expansion, those of order $A^0$, certainly 
{\em can\/} occur.  Their main effect in Eqn.\ (\ref{B3}) will be 
to shift the coefficient $1/4\hbar G$ in the Bekenstein-Hawking 
formula.  But at least for corrections induced by ordinary matter fields, 
such shifts can be shown to merely correspond to renormalizations of 
Newton's constant \cite{Fursaev}.  If this is the case in general---as, 
once again, one would expect from the semiclassical derivations of the 
Bekenstein-Hawking formula---then the first observable effects of
quantum corrections to the central charge will come at order $A^{-1}$ 
in $\delta c/c_{\hbox{\scriptsize\it class}}$.

Such corrections will, for large $A$, give area-independent terms in 
the entropy of the sort discussed in Refs.\ \cite{MSW,Cardosa}.  As
noted above, these are certainly important as independent tests of 
statistical mechanical formulations.  But for large $A$, they will
normally be dominated by the logarithmic corrections to the Cardy 
formula discussed in this paper.

\end{document}